\documentclass[a4paper,11pt]{article}
\usepackage{pos}
\usepackage{verbatim}

\title{The role of electron capture decay in the precision era of Galactic cosmic-ray data}
\ShortTitle{The role of EC decay in the precision era of GCR data}

\author*[a]{M. Borchiellini}
\author[b]{D. Maurin}
\author[a]{M. Vecchi}

\affiliation[a]{Kapteyn Astronomical Institute, University of Groningen\\
Landleven 12, 9747 AD Groningen, The Netherlands}

\affiliation[b]{LPSC, Université Grenoble Alpes, CNRS/IN2P3,\\
53 avenue des Martyrs, 38026 Grenoble, France}

\emailAdd{m.borchiellini@rug.nl}

\abstract{Electron capture (EC) decay relies on attachment and stripping cross-sections, that in turn, depend on the atomic number of the nucleus. We revisit the impact of EC decay in the context of the high-precision cosmic-ray fluxes measured by the AMS-02 experiment. We derive the solution of the steady-state fluxes in a 1D thin disk model including EC decay. We compare our results with relevant elemental and isotopic fluxes and evaluate the impact of this process, given the precision of recent AMS-02, ACE-CRIS, SuperTIGER, and Voyager data. We find this impact to be at the level or larger than the precision of recently collected data for several species, e.g. $_{31}$Ga and $_{33}$As, indicating that EC decay must be properly taken into account in the calculation.
}

\ConferenceLogo{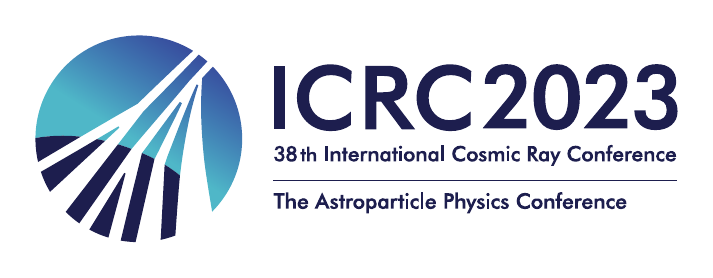}

\FullConference{%
38th International Cosmic Ray Conference (ICRC2023)\\
  26 July - 3 August, 2023\\
  Nagoya, Japan}

\begin{document}
\maketitle

\section{Introduction}
The study of  Galactic Cosmic Rays (GCRs) can provide information not only on their propagation and the properties of their sources but also on new physics phenomena in the Universe (e.g. dark matter).
Direct measurements currently provide high-precision data on GCR fluxes and the isotopic composition of heavy elements: AMS-02 measured GCR top-of-atmosphere (TOA) fluxes from H up to Si and Fe, at $\sim2~{\rm GV}-2$~TV, with unprecedented precision \cite{AMS:2021nhj},  SuperTIGER released TOA elemental ratios at 3.1~GeV/n for $26\leq Z\leq40$ \cite{Murphy:2016kyv}, whereas Voyager published interstellar (IS) fluxes at $\sim50-200$~MeV/n for H to Ni \cite{2016ApJ...831...18C}; 
ACE-CRIS also recently extended the measurements of the TOA isotopic composition at a few hundred of MeV/n for elements $29<Z<38$ \cite{Binns:2022qmw}.

For this reason, it becomes increasingly important to model the processes that contribute to GCR transport as accurately as possible to obtain models for GCR fluxes precise enough to compare them with the available data and search for new (astro)physics phenomena. One process that has not been often discussed in the literature is electron capture (EC) decay, which has been interpreted in the context of the leaky-box model in \cite{1985Ap&SS.114..365L}. It consists of the decay of a nuclide after the capture of a K-shell electron. Hence it does not occur freely in the Interstellar Medium (ISM), since GCRs are usually fully ionised. This implies that the effectiveness of EC decay depends heavily on the cross sections for attachment and stripping of electrons for the different GCR nuclei, hence on their atomic numbers, but also on their decay time which ranges from ms to Myr. In particular, a higher impact of EC decay is expected for heavy GCRs
\cite{1984ApJS...56..369L}.

This work aims to assess the impact of EC decay in the context of the high-precision GCR elemental fluxes measured by the ACE-CRIS, AMS-02, SuperTIGER, and Voyager experiment, and the isotopic ratios measured by ACE-CRIS. In Sect.~2, we discuss the general framework for GCR transport and the methods used in this analysis. In Sect.~3, we present our results, while in Sect.~4, we summarise our findings.

\section{Methodology}
The transport of GCRs is described by a diffusion-advection equation which has been extensively discussed in \cite{Genolini2019}. The differential density $n_{\alpha}$ of a GCR species $\alpha$ is given by
\begin{equation} 
\label{transport_gen}
\begin{split}
    - \vec{\nabla}_{\textbf{x}} \left\{ D(E) \vec{\nabla}_{\textbf{x}} n_{\alpha} - \vec{V}_c n_{\alpha}\right\} + & \frac{\partial}{\partial E} \left\{ b_{\rm tot}(E)n_{\alpha} - \beta^2 K_{pp} \frac{\partial n_{\alpha}}{\partial E}\right\} + \sigma_{\alpha}^{\rm inel} \, v_{\alpha} \, n_{\rm ISM} \, n_{\alpha} \, +\Gamma_{\alpha} \, n_{\alpha}\, \\
     & = q_{\alpha}\, + \sum_{\beta>\alpha}\left\{ \sigma_{\beta \rightarrow \alpha} \, v_{\beta} \, n_{\rm ISM}\, + \Gamma_{\beta \rightarrow \alpha} \right\} n_{\beta} \,,
\end{split}
\end{equation}
where the source term (right-hand side of the equation) is given by a primary injection rate $q_{\alpha}$, and a secondary injection rate from inelastic interactions of heavier species $\beta$ on the ISM (production cross-section $\sigma_{\beta \rightarrow \alpha}$) or from nuclear decay (rate $\Gamma_{\beta \rightarrow \alpha}$).
The other terms are, respectively, from left to right: 
the diffusion coefficient $D$ describing the scattering of CRs off magnetic turbulence, which depends on the rigidity $R=pc/Ze$; the galactic wind $V_c$; the rate for energy losses $b_{\rm tot}(E) \equiv dE/dt$ that includes ionisation and Coulomb processes as well as adiabatic losses induced by convection;
the energy-dependent coefficient $K_{pp}$ used to model reacceleration; the rate of inelastic interactions on gas $ \sigma_{\alpha}^{\rm inel} \,v_{\alpha} \, n_{\rm ISM}$ and the nuclear decay rate $\Gamma_{\alpha}$.

In this work,  we incorporate in Eq.~(\ref{transport_gen}) electron capture nuclides by treating the different charge states separately, following  \cite{1984ApJS...56..369L}. In this preliminary analysis, to study the interplay between the different processes analytically, we neglect energy losses, and we also neglect convection for simplicity. 
We perform our calculations in the two-zone (thin disk/thick halo) 1D propagation model, as used, for instance, in \cite{Genolini2019}, in which GCR fluxes only depend on the vertical coordinate $z$. The ISM gas (with density $n_{\rm ISM}$) and astrophysical sources are localised in a thin disc of half-height $h = 0.1$ kpc, and the thin disc is embedded in a thick halo, where GCR are confined, and they diffuse by scattering on magnetic fields irregularities. The halo is modelled as a slab in the radial direction with half-height $L = 5$ kpc, and the observer is located at $z=0$. 
For practical calculations, we model the diffusion coefficient as a power law with breaks both at low and high rigidities \cite{Genolini2019}, and the parameter values are taken from the combined analysis \cite{Maurin2022} of AMS-02 Li/C, Be/C, and B/C data. 

In this geometry and with the above approximations, the steady-state transport equation for an EC-unstable species takes the form of the following system of two coupled equations:
\begin{equation}
\label{eq:transp_EC}
    \begin{cases}
        -D(E)\,\partial^2_z n_0 + 2h\delta(z)  \biggr\{\Gamma^{\rm inel} n_0 + \Gamma^a n_0 - \Gamma^{\rm strip} n_1 \biggr\} = 2h\delta(z)q\,; \\
        -D(E)\,\partial^2_z n_1 + 2h\delta(z)  \biggr\{\Gamma^{\rm inel} n_1 + \Gamma^{\rm strip} n_1 - \Gamma^{\rm att} n_0  \biggr\} + \Gamma^{\rm EC}n_1 = 0\,.
    \end{cases}
\end{equation}
These two equations describe the spatial and energy evolution of the differential density of the fully ionised GCR ($n_0$) and the same GCR with one electron attached ($n_1$). The transition from one state to another is described by the electron stripping and attachment rates, denoted $\Gamma^{\rm strip} = n_{\rm ISM}\,  v \, \sigma_{\rm strip}$ and $\Gamma^{\rm att} = n_{\rm ISM} \,v \,\sigma_{\rm att}$ respectively; we take the cross-section parametrisations $\sigma_{\rm att}$ and $\sigma_{\rm strip}$ from \cite{1984ApJS...56..369L}. We assume here that higher charge states are almost not populated \cite{1984ApJS...56..369L}, and in order to have a close system (constant number density of the species considered), we do not allow $n_1$ to attach electrons. As EC-unstable species decay by capturing a K-shell electron, the EC decay rate, $\Gamma^{\rm EC} = 1/(\tau_{\rm EC} \, \gamma) $, is implemented for $n_1$ only.
\renewcommand{\arraystretch}{1.8}
\setlength{\tabcolsep}{10pt}
\begin{table}[b]
\centering
\begin{tabular}{lll}
Process              & Timescale & Dependencies \\ \hline
Diffusion            &   $t_D = \dfrac{L^2}{2D}$  & $ D \propto E^{0.5}$ \\ 
Inelastic scattering & $t_{\rm inel} = \dfrac{1}{n_{\rm ISM}\,  v \, \sigma_{\rm inel}}$  &      $\sigma_{\rm inel} \propto A^{2/3}$             \\ 
Attachment           &  $t_{\rm att} = \dfrac{1}{n_{\rm ISM}\,  v \, \sigma_{\rm att}}$         &    $\sigma_{\rm att} \propto \sigma(E)Z^2$               \\ 
Stripping            &  $t_{\rm strip} = \dfrac{1}{n_{\rm ISM}\,  v \, \sigma_{\rm strip}}$         &  $\sigma_{\rm strip} \propto \sigma(E)Z^{-2}$                 \\ 
EC decay             & $t_{\rm EC} = \tau_{\rm EC} \, \gamma$          &         $t_{\rm EC} \propto E$          \\ \hline
\end{tabular}
\caption{The five competing processes that have been taken into account to model GCR  propagation, with their corresponding timescales and dependencies.}
\label{tab:timescales}
\end{table} 
\renewcommand{\arraystretch}{1}
\setlength{\tabcolsep}{6pt}

\section{Results}
\subsection{Timescales}

Before showing the solutions of Eq.~(\ref{eq:transp_EC}), it is interesting to discuss the timescales in our 1D model since their interplay affects the final isotopic and elemental fluxes.
Five propagation processes have been considered, as reported in Table~\ref{tab:timescales}, where we highlight the main dependencies on energy or atomic number.
\begin{figure}[b]
\includegraphics[width=12cm]{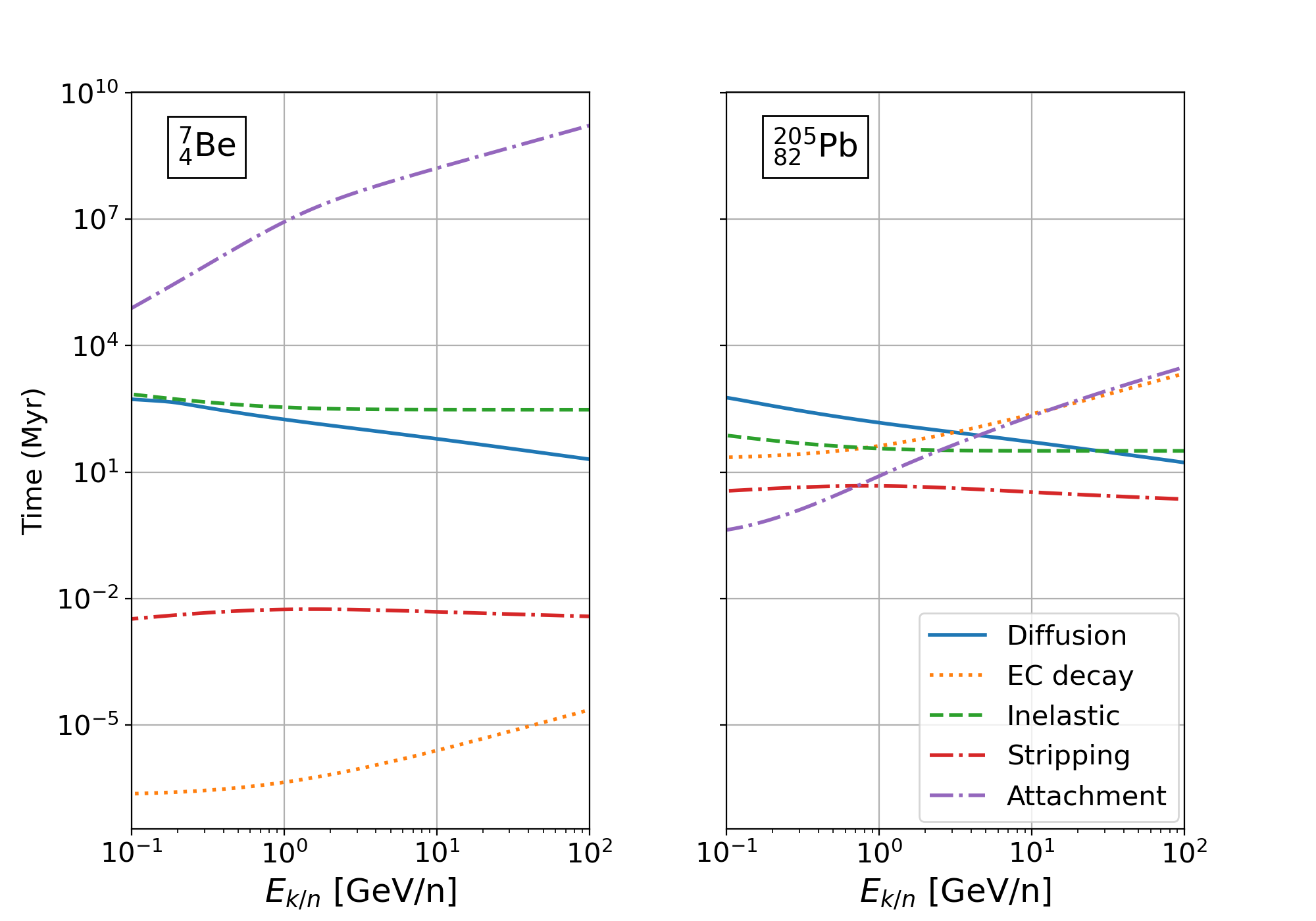}
\centering
\caption{Timescales for the processes listed in Table~\ref{tab:timescales} as a function of kinetic energy per nucleon, for the species $_4^{7}$Be with $t_{1/2}=1.46\;10^{-7}$~Myr (left panel) and $^{205}_{82}$Pb with $t_{1/2}=1.4\;10^7$~Myr (right panel).}
\label{timescales}
\end{figure}
The associated timescales are shown in Fig.~\ref{timescales} as a function of the kinetic energy per nucleon, for $_4^{7}$Be (left panel) and $^{205}_{82}$Pb (right panel); these two species are representative of light and heavy EC decaying nuclides respectively.

First, we recover the standard result (e.g. \cite{2019A&A...627A.158D}) that diffusion dominates at high energy (smallest timescale), while inelastic scattering is relevant mostly at low energies, especially for heavy nuclei.
Then, for EC decay to dominate over the other propagation processes, a first condition is that $t_{\rm EC}$ (orange dotted lines) has to be lower than $t_D$ (blue solid lines) and $t_{\rm inel}$ (green dashed lines), which is always more likely to happen at low energy, as $t_{\rm EC}\propto E$, while $t_{D} \propto 1/\sqrt{E}$ and $t_{\rm inel}$ is roughly constant. However, the net effect of EC decay also relies on the interplay between attachment (magenta dash-dotted lines) and stripping processes (red dash-dotted lines), which depend on the kinetic energy and the atomic number through their cross-sections: as seen from Fig.~\ref{timescales}, attachment only overcomes stripping for low energy and heavy nuclei, as $t_{\rm att}\propto Z^2$ while $t_{\rm strip}\propto Z^{-2}$. As a result, the impact of EC decay will depend on the specific ordering of these three times, and a species will disappear via EC decay only if both $t_{\rm att} \lesssim t_{D}$ and  $t_{\rm EC}\lesssim(t_{D},t_{\rm strip})$.   

\subsection{Impact of EC decay on isotopic and elemental fluxes}
We solve the coupled system of Eq.~(\ref{eq:transp_EC}) following \cite{Maurin2001}, and we obtain for the differential density at $z=0$:

\begin{equation} \label{eq:solutions}
    \begin{cases}
         n_1 = \dfrac{\Gamma^{\rm att} n_0}{\sqrt{\dfrac{D(E)\, \Gamma^{\rm EC}}{h^2}}\,\coth\left(\sqrt{\dfrac{\Gamma^{EC}}{D(E)}}L\right) + \Gamma^{\rm inel} + \Gamma^{\rm strip}}\;; \\
         n_0 = \dfrac{q}{\dfrac{D(E)}{hL} + \Gamma^{\rm inel} + \Gamma^{\rm att} - \Gamma^{\rm strip} \Gamma^{\rm att} \left[\sqrt{\dfrac{D(E)\, \Gamma^{\rm EC}}{h^2}}\,\coth\left(\sqrt{\dfrac{\Gamma^{\rm EC}}{D(E)}}L\right) + \Gamma^{\rm inel} + \Gamma^{\rm strip}\right]^{-1}}\;.
    \end{cases}
\end{equation}

\begin{figure}[b]
\includegraphics[width=12cm]{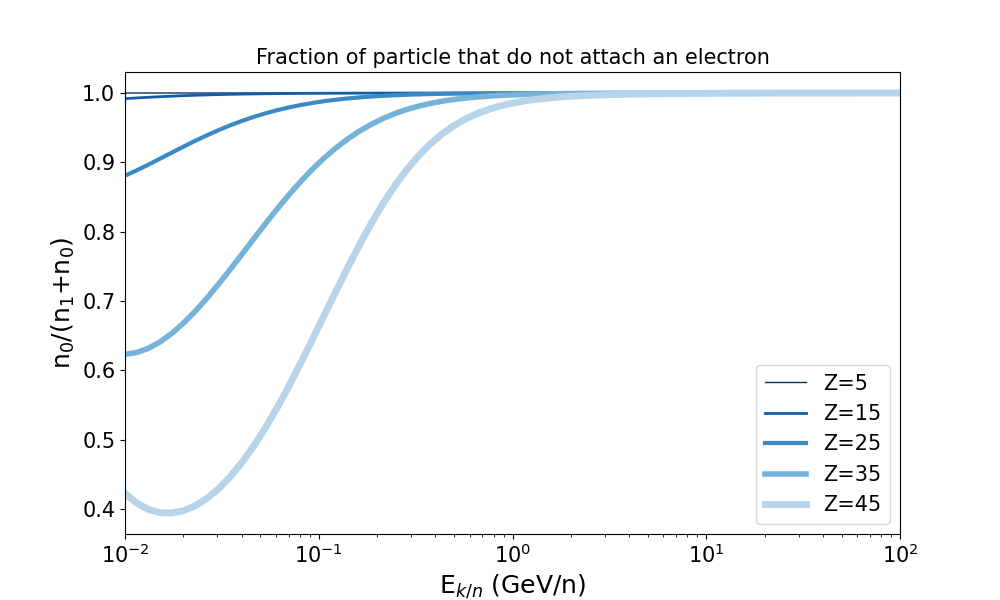}
\centering
\caption{Fraction $n_0/(n_0 + n_1)$ of GCRs that do not attach an electron, where $n_0$ and $n_1$ are defined in Eq.~(\ref{eq:solutions}). Different values of $Z$ are shown as different shades of blue and line thicknesses.}
\label{stable_ab}
\end{figure}
Since the balance between attachment and stripping plays such a critical role in the effectiveness of EC decay, in Fig.~\ref{stable_ab} we examine, disregarding EC decay (i.e. considering $\tau_{\rm EC}\to\infty$ in Eq.~\ref{eq:solutions}), the fraction of GCRs that do not attach an electron ($n_0$) with respect to the total number density ($n_0+n_1$), for growing elements $Z$ (from thin to thick lines). The above conclusions from the study of characteristic timescales can explain the trend shown by the different lines: no electrons are attached above a few GeV/n, coherently with a scenario in which $t_{\rm att} \gg t_D$; secondly, heavier GCRs attach more electrons than light GCRs due to the interplay between stripping and attachment cross sections. Overall, the fraction of attached electrons is at most $\gtrsim  0.5$ for $Z\leq40$ at $E_{k/n}\sim 10$~MeV/n.

Taking into account the half-life of EC-unstable species, we can now evaluate {the impact of EC decay on the relevant isotopes and associated elements. We selected a subset of species $Z\leq 40$ with both short and intermediate half-lives, which are listed in Table~\ref{tab:isotopes}.
These values have been used to compute the final isotopic and elemental abundances and derive the results presented in Fig.~\ref{impact_is}.
The top panel of Fig.~\ref{impact_is} shows the percentage of GCR isotopes that decay by EC. Unsurprisingly, EC decay has no impact on isotopic fluxes above a few GeV/n per nucleon. At lower $E_{k/n}$, there is almost no visible effect for intermediate-lived isotopes (orange dashed lines), while short-lived GCRs (solid blue lines) exhibit different behaviours depending on their atomic number. In particular, the heavier nuclei ($^{67}_{31}$Ga and $^{73}_{33}$As) decay almost completely below 100~MeV/n. 

\begin{table}[h]
    \centering
    \begin{tabular}{ccc}
    \hline 
    \rule{0pt}{4ex}   Isotope & $t_{1/2}$ (Myr) & Isotopic fraction\\
    \hline
    \rule{0pt}{4ex}  $^7_4$Be & 1.46 $10^{-7}$ & 0.55 \\[3pt]
    $^{37}_{18}$Ar & 9.58 $10^{-8}$ & 0.30 \\[3pt]
    $^{41}_{20}$Ca & 1.00 $10^{-1}$ & 0.07 \\[3pt]
    $^{44}_{22}$Ti & 4.70 $10^{-5}$ & 0.04 \\[3pt]
    $^{53}_{25}$Mn & 3.70 & 0.35 \\[3pt]
    $^{67}_{31}$Ga & 8.93 $10^{-9}$ & 0.07\\[3pt]
    $^{73}_{33}$As & 2.20 $10^{-7}$ & 0.36 \\[3pt]
    \hline
    \end{tabular}
    \caption{Sample of EC-unstable GCRs $Z\leq40$, with their EC half lives \cite{1984ApJS...56..369L} and GCR isotopic fractions at low energy (from data extracted from the CR Data Base \cite{Maurin:2023alp}).}
\label{tab:isotopes}
\end{table}
\begin{figure}[t]
\centering
\includegraphics[width=12cm]{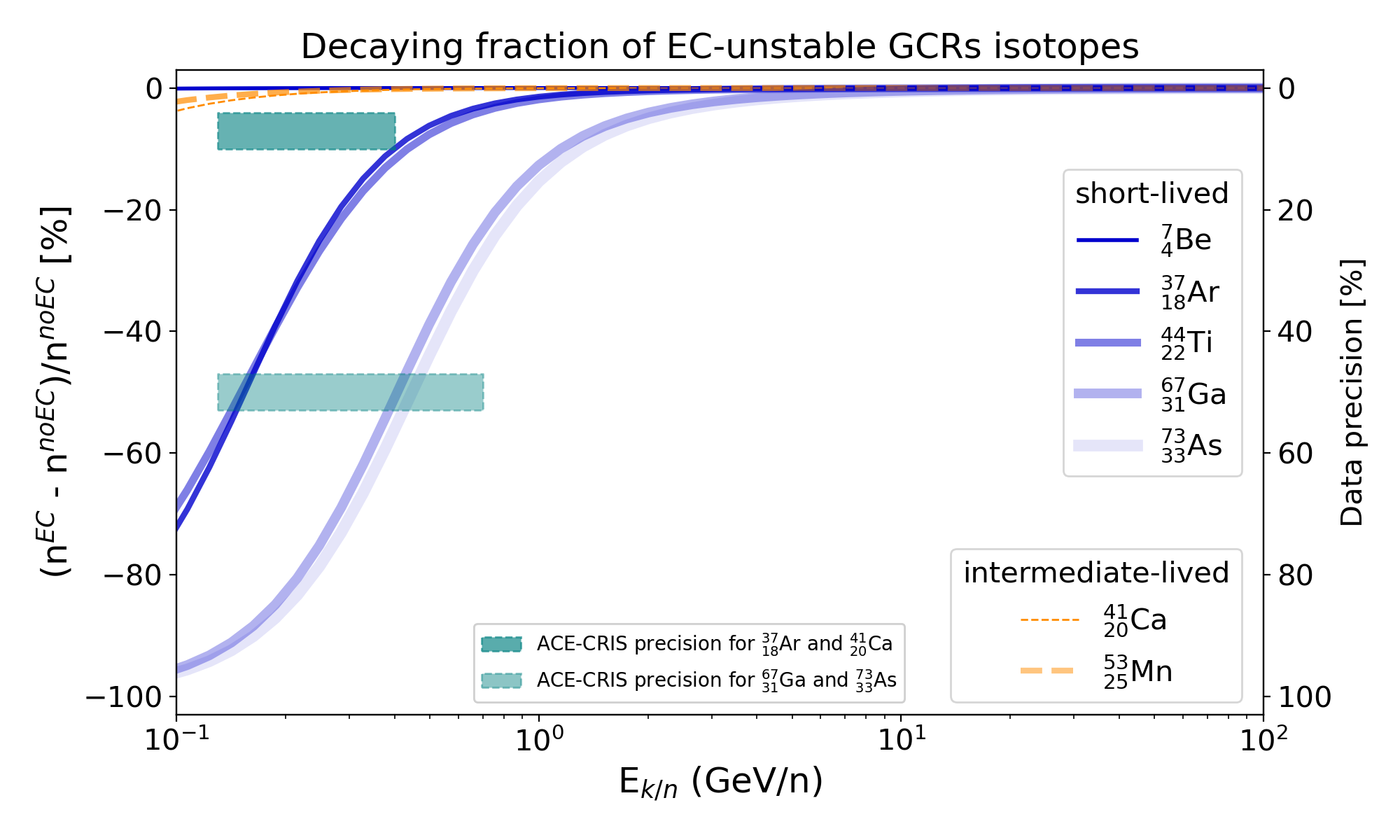}
\includegraphics[width=12cm]{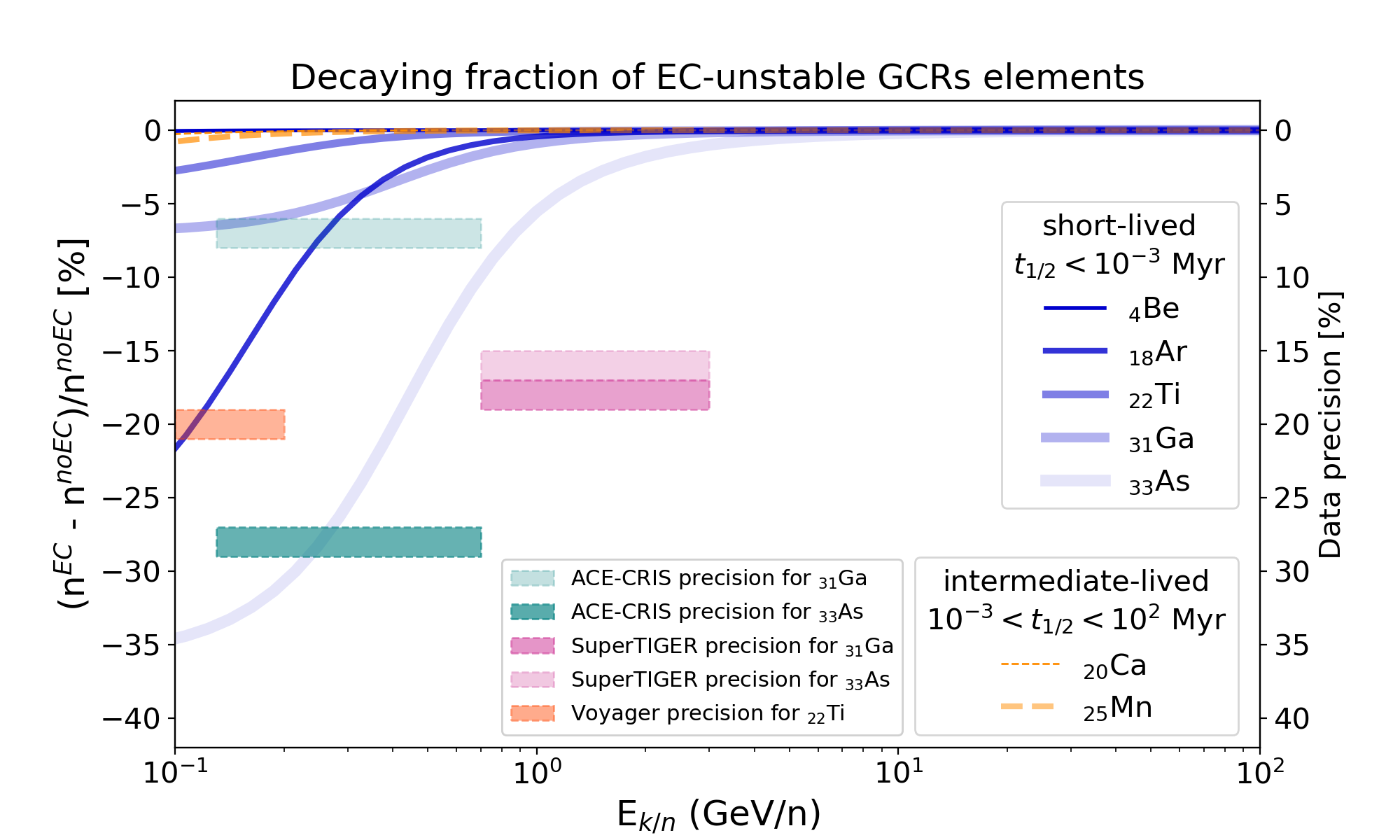}
\caption{Decaying fraction of a sample of EC-unstable species in GCR isotopic fluxes (top) and associated elemental fluxes (bottom) computed as $(n^{\rm EC}-n^{\rm noEC})/n^{\rm noEC}$, where $n=n_0+n_1$ from Eq.~(\ref{eq:solutions}) with ($n^{\rm EC}$) and without ($n^{\rm noEC}$) the EC decay term. In the bottom panel, the abundances of the different isotopes have been weighted by their isotopic GCR fractions. The shaded areas correspond to the precision (right-hand side axis) and energy range of recent experimental data.}
\label{impact_is}
\end{figure}
It is interesting to compare the impact of EC decay (observed in our simplified model) to the precision of recent data --- we recall that the flux $J$ is related to the differential density $n$ by $J = vn/(4\pi)$, so that the relative differences (considered below) on $n$ and $J$ are one and the same.
Experimentally, light nuclei are more abundant than heavier ones, with a strong suppression of elements heavier than Fe. For this reason, light isotopes have been measured with better precision. However, light EC-unstable isotopes are rare, and because of the large attachment time for light nuclei, the abundance of $^7_4$Be (thinnest blue line in the top panel of Fig.~\ref{impact_is}) does not show any change with respect to a model without EC decay. Abundances for GCR isotopes in the range $Z=15-40$ have been measured by the ACE-CRIS experiment \cite{Binns:2022qmw,2009ApJ...695..666O}. Their precision is dominated by statistical uncertainties and strongly isotope-dependent: at a few hundreds of MeV/n it has a typical value $\lesssim10\%$ for $Z=15-30$, reaching a precision $\lesssim 50\%$ for $Z=30-40$. We predict the impact of EC decay on $^{37}_{18}$Ar flux to be $\geq 10\%$ for $E_{k/n} \lesssim 400$~MeV/n, which is higher than ACE-CRIS precision for the same isotope and energy range. On the other hand, the precision for ACE-CRIS on $^{67}_{31}$Ga and $^{73}_{33}$As for at a few MeV/n is $\sim 50\%$, of the order of the impact of EC decay on the modelled fluxes at the same energies (in practice, Solar modulation shifts data TOA energy towards higher IS ones, i.e. energies with even smaller EC impact in our IS calculations).

The impact of EC-decay on elemental fluxes is shown in the bottom panel of Fig.~\ref{impact_is}. This impact is calculated by assuming EC-unstable species constitute a fraction of the elemental flux. In practice, we set this isotopic fraction to a constant value with energy based on the GCR measured one; the associated numbers are reported in Table~\ref{tab:timescales}. The impact of EC-decay is thus diluted in the elemental fluxes, but the latter are easier to measure than isotopic ones due to intrinsic experimental challenges in isotopic separation. AMS-02 has already published the elemental flux of all species from H to S and Fe, and the flux of He isotopes, with a precision reaching at best a few percent~\cite{AMS:2021nhj}, for energies typically $\gtrsim 500$~MeV/n. At these energies, the impact on $_{18}$Ar is slightly larger than the expected AMS-02 data precision for its flux.
The precision for $_{22}$Ti Voyager data is $\sim 20\%$ between $100$ and $200$~MeV/n, of the order of the impact of EC decay on the same elemental flux in that energy range.
The impact of EC decay on $_{31}$Ga and $_{33}$As fluxes for $E_{k/n} \sim 100$~MeV/n correspond to the value of the whole isotopic fraction ($\sim 7\%$ and $\sim 36\%$, respectively), since the corresponding EC-unstable isotopes fully decay at low energies per nucleon. 
SuperTIGER precision, on the other hand, is $\sim 16\%$ for $_{31}$Ga and $\sim 18\%$ for $_{33}$As at $E_{k/n}$ above $700$~MeV/n, where the impact of EC decay is already strongly suppressed. 
The precision of ACE-CRIS measurements for $_{31}$Ga and $_{33}$As abundances at a few hundreds of MeV/n is at most of the order of the impact of EC decay on the same elemental fluxes, with values of $\sim 7\%$ and  $\sim 28\%$ for $_{31}$Ga and $_{33}$As respectively.

\section{Conclusions}
In the context of recent high-precision data, we have revisited the impact of EC-decay on GCR fluxes. In a 1D diffusion model with parameters tuned to recent secondary-to-primary data, we found that EC decay impacts isotopic fluxes at most at the level of $\lesssim 50\%$ at a few hundreds of MeV/n, and $\lesssim 20\%$ for elemental fluxes in the same energy range. 

These numbers are of the order of ACE-CRIS precision for isotopic fluxes and slightly larger than AMS-02 precision for elemental fluxes. The impact of EC decay at very low energies is of the same order of Voyager precision for $_{22}$Ti flux, while at energies higher than $700$ MeV/n it is lower than SuperTIGER precision for elemental abundances. Overall,  this shows that this effect has to be taken properly into account in the calculation.

The analysis presented here will be improved in several directions. First, the analytical solution can be further exploited to assess whether the attachment of several electrons needs to be taken into account. In particular, as $Z\gtrsim 30$ data have so far been interpreted in a leaky-box model only, it is important to compare the impact of EC decay in the leaky-box and in a more realistic diffusion model. To do so, energy losses, Solar modulation, and the full source terms and fragmentation terms need to be accounted for, and we are implementing species $Z>30$ in the USINE code~\cite{Maurin:2018rmm}.

\end{document}